\documentclass[10pt,conference]{IEEEtran}
\IEEEoverridecommandlockouts
\usepackage{amsmath,amssymb,amsfonts}                           
\usepackage{hyperref}                                           
\usepackage[caption=false]{subfig}                              
\usepackage[linesnumbered,algoruled,boxed,lined]{algorithm2e}   
\usepackage[dvipsnames]{xcolor}
\usepackage{tikz}                                               
\usetikzlibrary{quantikz2}                                      
\usepackage{blindtext}
\usepackage{cleveref}
\usepackage{graphicx}
\usepackage{orcidlink}
\usepackage{etoolbox}
\makeatletter
\patchcmd{\@makecaption}
  {\scshape}
  {}
  {}
  {}
\makeatletter
\patchcmd{\@makecaption}
  {\\}
  {.\ }
  {}
  {}
\makeatother
\def\tablename{Tab.}

\usepackage{bbold}
\usepackage{multirow}
\usepackage{upgreek}
\usepackage{cite}
\usepackage[normalem]{ulem}

\usepackage{amsthm}
\usepackage{mathtools}

\DeclareMathOperator*{\argmin}{arg\,min}
\usepackage{braket}
\theoremstyle{plain}
\newtheorem{theorem}{Theorem}

\newtheorem{definition}{Definition}
\theoremstyle{definition}

\DeclareMathOperator{\polylog}{polylog}
\tikzset{%
    uctrl/.style={draw, circle, minimum size=0.285pc, append after command={ \pgfextra { \fill  (0,0)-- (225:2pt) arc (225:405:2pt) -- cycle; } } }
}
\DeclareExpandableDocumentCommand{\uctrl}{O{}{m}}{|[uctrl,#1]| {#2} \qw}

\tikzset{%
    pctrl/.style={draw, circle, minimum size=0pc, append after command={ \pgfextra { \path[draw=black,fill=white]   (-0.075,0.055)-- (0.075,0.055) -- (0,-0.095) -- (-0.075,0.055) -- cycle; } } }
}
\DeclareExpandableDocumentCommand{\pctrl}{O{}{m}}{|[pctrl,#1]| {#2} \qw}

\tikzset{%
    zctrl/.style={draw, circle, minimum size=0.285pc, fill=white, append after command={ \pgfextra { \path[draw=black,fill=black]  (0:0.2pt) arc (0:360:0.2pt) -- cycle; } } }
}
\DeclareExpandableDocumentCommand{\zctrl}{O{}{m}}{|[zctrl,#1]| {#2} \qw}

\newsavebox{\boxQ}
\AtBeginDocument{
    \newwrite\bibnotes
    \def\bibnotesext{Notes.bib}
    \immediate\openout\bibnotes=\jobname\bibnotesext
    \immediate\write\bibnotes{@CONTROL{REVTEX42Control}}
    \immediate\write\bibnotes{@CONTROL{
    apsrev42Control,author="48",editor="1",pages="1",title="0",year="1"}}
     \if@filesw
     \immediate\write\@auxout{\string\citation{apsrev42Control}}
    \fi
}

\usetikzlibrary{chains}

\makeatletter
\DeclareRobustCommand{\rvdots}{%
  \vbox{
    \baselineskip4\p@\lineskiplimit\z@
    \kern-\p@
    \hbox{.}\hbox{.}\hbox{.}
  }}
\makeatother

\makeatletter
\DeclareRobustCommand{\rvdotstwo}{%
  \vbox{
    \vspace{3pt}
    \baselineskip3\p@\lineskiplimit\z@
    \kern-\p@
    \hbox{.}\hbox{.}\hbox{.}
    \vspace{3pt}
  }}
\makeatother

\makeatletter
\DeclareRobustCommand
  \rddots{\mathinner{\mkern1mu\raise7\p@
    \vbox{\hbox{.}}\mkern2mu
    \raise4\p@\hbox{.}\mkern2mu\raise\p@\hbox{.}\mkern1mu}}
\makeatother

\makeatletter
\DeclareRobustCommand
  \lddots{\mathinner{
  \mkern2mu\raise\p@\hbox{.}
  \mkern2mu\raise4\p@\hbox{.}
  \mkern1mu\raise7\p@\vbox{\hbox{.}}\mkern1mu}}
\makeatother

\makeatletter
\def\thickhline{%
  \noalign{\ifnum0=`}\fi\hrule \@height \thickarrayrulewidth \futurelet
   \reserved@a\@xthickhline}
\def\@xthickhline{\ifx\reserved@a\thickhline
               \vskip\doublerulesep
               \vskip-\thickarrayrulewidth
             \fi
      \ifnum0=`{\fi}}
\makeatother

\newlength{\thickarrayrulewidth}
\makeatletter
\renewcommand{\qwbundle}[2][]{
  \pgfkeys{/quantikz/gates/.cd,style=,Strike Width=0.08cm,Strike Height=0.12cm,#1}%
  \pgfkeysgetvalue{/quantikz/gates/style}{\qz@style}%
  \pgfkeysgetvalue{/quantikz/gates/Strike Width}{\qz@sw}%
  \pgfkeysgetvalue{/quantikz/gates/Strike Height}{\qz@sh}%
  \expanded{%
    \noexpand\arrow[strike arrow={\qz@sw}{\qz@sh}{\unexpanded{#2}},\qz@style,phantom]{l}%
  }%
}
\makeatother

\makeatletter
\DeclareRobustCommand{\pbar}{\mathord{%
  \text{$\m@th\mkern-2mu\raisebox{-1.5ex}[0pt][0pt]{$\mathchar'26$}\mkern-7mu p$}%
}}
\makeatother

\makeatletter
\DeclareRobustCommand{\Bbar}{\mathord{%
  \text{$\m@th\mkern-2mu\raisebox{-0.85ex}[0pt][0pt]{$\mathchar'26$}\mkern-9mu B$}%
}}
\makeatother

\makeatletter
\DeclareRobustCommand{\Bbarc}{\mathord{%
  \text{$\m@th\mkern-2mu\raisebox{-0.85ex}[0pt][0pt]{$\mathchar'26$}\mkern-8mu \mathcal{B}$}%
}}
\makeatother

\makeatletter
\DeclareRobustCommand{\Lbarc}{\mathord{%
  \text{$\m@th\mkern-2mu\raisebox{-0.6ex}[0pt][0pt]{$\mathchar'26$}\mkern-9mu \mathcal{L}$}%
}}
\makeatother

\makeatletter
\DeclareRobustCommand{\thetabar}{\mathord{%
   \text{$\m@th\mkern-2mu\raisebox{-0.95ex}[0pt][0pt]{$\mathchar'26$}\mkern-5.1mu  \theta$}%
}}
\makeatother

\def\linkurl#1{\url{#1}}

\begin{document}

\title{End-to-End Speedup for Quantum Simulation-Based Optimization in Power Grid Management}

\makeatletter
\newcommand{\linebreakand}{%
  \end{@IEEEauthorhalign}
  \hfill\mbox{}\par
  \mbox{}\hfill\begin{@IEEEauthorhalign}
}
\makeatother

\DeclareRobustCommand*{\IEEEauthorrefmark}[1]{%
  \raisebox{0pt}[0pt][0pt]{\textsuperscript{\footnotesize\ensuremath{#1}}}}

\renewcommand\footnoterule{\vspace*{-3pt}\hrule width 1in\vspace*{2.6pt}}

\author{
\IEEEauthorblockN{%
Jonas Stein\textsuperscript{\orcidlink{0000-0001-5727-9151}, }\thanks{\textsuperscript{$\dagger$}Contact to corresponding author: \href{mailto:jonas.stein@ifi.lmu.de}{jonas.stein@ifi.lmu.de}.}\IEEEauthorrefmark{1,2}\textsuperscript{, $\dagger$}, Jannis Lutz\IEEEauthorrefmark{1}, Moritz Sölderer\IEEEauthorrefmark{1}, Maximilian Adler\textsuperscript{\orcidlink{0009-0003-1101-3069}, }\IEEEauthorrefmark{2},
\\ Michael Lachner\textsuperscript{\orcidlink{0009-0008-6874-8329}, }\IEEEauthorrefmark{2}, David Bucher\textsuperscript{\orcidlink{0009-0002-0764-9606}, }\IEEEauthorrefmark{2}, and Claudia Linnhoff Popien\textsuperscript{\orcidlink{0000-0001-6284-9286}, }\IEEEauthorrefmark{1}}
\IEEEauthorblockA{\IEEEauthorrefmark{1}\textit{QAR-Lab, Department of Computer Science, \href{https://ror.org/05591te55}{LMU Munich}, Munich, Germany}\\\IEEEauthorrefmark{2}\textit{Aqarios GmbH, Munich, Germany}
}}

\maketitle

\bstctlcite{BSTcontrol}

\begin{abstract}
Quantum Simulation-based Optimization (QuSO) is a recently proposed class of optimization problems that entails industrially relevant problems characterized by cost functions or constraints that depend on summary statistic information about the simulation of a physical system or process. This work extends initial theoretical results that proved an up-to-exponential speedup for the simulation component of the QAOA-based QuSO solver for the unit commitment problem \cite{stein2024exponentialquantumspeedupsimulationbased} to an end-to-end speedup, explicitly including the outer optimization component. The numerical experiments were conducted using randomly generated power grid instances of varying sizes and loads that adhere to the physical properties of real world power grids. Exploiting clever classical pre-computation, we develop a very efficient classical quantum circuit simulation that bypasses costly ancillary qubit requirements of the original algorithm, allowing for large-scale experiments. We show that 16 QAOA layers suffice to outperform a strong classical baseline for problems involving up to 14 qubits in scenarios of high load and perform on par otherwise. In summary, our results thus extend previous partial quantum speedup results for QuSO problems to an end-to-end setting that encompasses the runtime of the complete algorithm for a problem of industrial relevance.

\begin{IEEEkeywords}
Benchmarking, Simulation-Based Optimization, QAOA, Unit Commitment.
\end{IEEEkeywords}
\end{abstract}

\section{Introduction}
\label{sec:introduction}

Quantum computing has the potential to significantly impact both optimization and simulation tasks, with theoretical results indicating possible exponential improvements in certain settings~\cite{HHL,Ambainis12,Berry15,Childs17,gilyen2018quantum,SSO19,Lin2020optimalpolynomial,posDefQLSP,montanaro2024quantumspeedupssolvingnearsymmetric}. 
A recently introduced class of problems, called Quantum Simulation-based Optimization (QuSO), deals with scenarios in which evaluating candidate solution requires complex simulations, such as solving systems of linear equations or partial differential equations~\cite{stein2024exponentialquantumspeedupsimulationbased}. In these problems, the simulation outcome is used either to compute the objective value or to verify feasibility constraints. A simple example of such a problem can take the form of
\begin{equation}
    \argmin_{x\in\{0, 1\}^n} f(x,y) \quad \textnormal{s.t.} \quad A y = b(x),
\end{equation}
for an arbitrary $y\in\mathbb{R}^m$, given fixed $b:x\mapsto \mathbb{R}^m$, and given fixed $A\in \mathbb{R}^{m\times m}$ invertible. For this example, the equality constraint $Ay=b(x)$ would constitute a simulation problem, and $f:(x,y)\mapsto \mathbb{R}$ may denote a representative of a class of $\mathsf{NP}$-hard polynomial objective functions. As $m$ is often very large, the predominant classical approach to solve such problems is iterative search, i.e., (1) probing an initial solution $x\in\{0, 1\}^n$, (2) solving the system of linear equations $Ay=b(x)$ to yield the value of $y$ allowing for (3), the computation of the objective value $f(x,y)$ and finally (4), trying to find better solutions by repeating these steps with a newly guessed solution candidate~\cite{10.1145/268437.268460}.

Problems of this type appear in a wide range of applications, such as pharmaceutical development \cite{myers2016response}, car and aircraft design \cite{Hoelscher_2026,balabanov1996topology}, or topology optimization \cite{Hoelscher_2026_2}, where detailed physical models must be optimized under various constraints. A defining characteristic of QuSO problems is that the information required from the result of the simulation must be easily extractable, i.e., it must take the form of \emph{summary statistic} information (cf. Ref.~\cite{HHL} and \Cref{def:SSI}).
This property makes it feasible to use quantum algorithms to perform the simulation component of the problem without encountering the full overhead of quantum state readout, thereby avoiding the so-called \emph{output problem}. This specific requirement distinguishes QuSO from traditional simulation-based optimization and allows it to achieve significant quantum speedups.
In Ref.~\cite{stein2024exponentialquantumspeedupsimulationbased}, Stein \textit{et al.} proposed a combination of the Quantum Approximate Optimization Algorithm (QAOA) to solve the optimization component with the Quantum Singular Value Transformation (QSVT) to compute the simulation outcome. They further demonstrated the potential for up-to-exponential quantum advantage in the context of the unit commitment problem, which is essentially the problem of optimizing the energy production costs by choosing which power generators to activate for a given power demand of all loads in a power grid. There, a simulation task arises inside the power transmission costs, which are determined by how much power flows over which transmission line, requiring a simulation of the overall power flow. For this provably QuSO-type problem, the proposed algorithm was shown to reduce the computational complexity of the simulation component to \( \tilde{\mathcal{O}}(\log(N) \kappa s) \), compared to the classical \( \tilde{\mathcal{O}}(N \sqrt{\kappa} s) \) complexity achieved by the conjugate gradient method~\cite{hestenes1952methods}, where $N$ denotes the problem size, $\kappa$ the condition number of the system of linear equations (SLE) to be solved (which is usually linear in $N$ for real-world applications) and $s$ the sparsity of this SLE. In our setting we adopt a slightly more general complexity characterization. Specifically, the classical baseline scales as $\mathcal{O}(N s \kappa \log(1/\epsilon))$ while the QAOA-based approach exhibits a complexity of $\mathcal{O}(\text{polylog}(N)\kappa^* s / (\epsilon \delta))$, where $\epsilon$ denotes the desired accuracy, $\delta$ the success probability, and $\kappa^*$ captures the effective condition number in the quantum setting. ~\cite{Shewchuck_ConjugateGradientMethods, stein2024exponentialquantumspeedupsimulationbased}

The main contribution of this paper is to investigate the performance of the optimization component of the QuSO solver proposed in Ref.~\cite{stein2024exponentialquantumspeedupsimulationbased}, as in that reference, only the simulation component was explored for possible quantum advantages. More concretely, we develop an efficient classical simulation of the quantum algorithm and perform an empirical evaluation of the proposed QuSO solver against a state-of-the-art classical baseline (i.e. simulated annealing). The experiments are conducted on randomly generated power grid instances with varying energy demand levels (loads). Details on the dataset generation process are provided in \Cref{sec:experimental_setup}.  
The classical circuit simulation algorithm is implemented by cleverly simplifying the proposed QuSO solver to a diagonalized QAOA using extensive classical precomputation of quantum subroutines based on ideas of Ref.~\cite{10821080}. Our results demonstrate that the quantum approach achieves competitive performance, attaining an average random adjusted approximation ratio (RAAR) of 69\% across all problem instances with up to 14 qubits -- which is comparable to results of the classical baseline. 

The remainder of this paper is structured as follows. We first outline the necessary preliminaries on the employed QuSO solver in \Cref{sec:background}. We then introduce the utilized problem formulation of the unit commitment problem and present our approach to solving it using a simplified version of the originally proposed QuSO solver for classical circuit simulation in \Cref{sec:methodology}. The experimental setup including baselines and datasets is outlined in \Cref{sec:experimental_setup}. The results of the empirical benchmark study are described in \Cref{sec:results} and discussed in \Cref{sec:discussion}. Finally, we conclude our findings in \Cref{sec:conclusion}.

\section{Quantum Simulation Based Optimization}
\label{sec:background}
This section outlines a formal definition of QuSO according to Ref.~\cite{stein2024exponentialquantumspeedupsimulationbased}, foundational concepts of quantum optimization via the Quantum Approximate Optimization Algorithm (QAOA), as well as the employed QuSO solver.
\subsection{Quantum Simulation-based Optimization}
QuSO is defined as a subset of Mixed-Integer Nonlinear Programming (MINLP), which represents a most extensive class of optimization problems. For simplicity, we focus solely on minimization problems in the remainder of this paper.

\begin{definition}
    A Mixed-Integer Nonlinear Programming (MINLP) problem consists of a function \( f(x) \) subject to constraints \( c_j(x) \leq 0 \) for all \( j \in [K] \), with continuous bounds \( x_i \in [l_i, u_i] \subset \mathbb{R} \) and integer constraints \( x_i \in \mathbb{Z} \) for certain indices \( i \in I \subseteq [n] \), where \( f : \mathbb{R}^n \rightarrow \mathbb{R} \) and \( c : \mathbb{R}^n \rightarrow \mathbb{R}^K \) are continuous functions. Solving the problem reduces to minimizing \(f\).
\end{definition} 
QuSO is defined based on the concept of summary statistic information, which we now define formally:

\begin{definition}[Summary Statistic Information, cf. Ref.~\cite{summary_statistic}]\label{def:SSI}
    Given an oracle \( O \) that prepares an \( n \)-qubit quantum state \( \ket{\psi} = O\ket{0}^{\otimes n} \), we define summary statistic information as the binary-encoded output from a quantum algorithm that relies on \( \ket{\psi} \), given access to \( O \).
\end{definition} 

\begin{definition}[Quantum Simulation-based Optimization (QuSO)] A QuSO problem is a MINLP problem where the objective function and/or constraints depend on the summary statistic result of a simulation problem, as in:
\[
\begin{aligned}
    & \text{minimize}_{x} && f(x, u(s(x))) \\
    & \text{subject to} && c_j(x, u(s(x))) \leq 0, \;\forall j \in [K], \\
    & && x_i \in [l_i, u_i] \subset \mathbb{R}, \\
    & && x_i \in \mathbb{Z}, \;\forall i \in I \subseteq [n],
\end{aligned}
\]
where \( f : \mathbb{R}^n \times \{0, 1\}^m \rightarrow \mathbb{R} \) and \( c : \mathbb{R}^n \times \{0, 1\}^m \rightarrow \mathbb{R}^K \) are continuous functions, \( s : \mathbb{R}^n \rightarrow \mathbb{R}^M \) represents a simulation problem, and \( u : \mathbb{R}^M \rightarrow \{0, 1\}^m \) extracts summary statistic information from \( s(x) \).
\end{definition} 
QuSO encompasses all simulation-based optimization problems that can potentially benefit from quantum speedup in the simulation component. While this paper focuses on simulations that can be natively expressed as a system of linear equations, QuSO also includes problems with non-linear simulations problems (e.g., Navier-Stokes equations~\cite{Gaitan2020}). QuSO’s main feature is its requirement that information from the simulation results must be extractable, which if can be done reasonably efficiently implies that obtaining summary statistics does not negate any quantum speedup from the simulation process.

\subsection{The Quantum Approximate Optimization Algorithm}
Being the basis of the employed QuSO solver, we now outline the Quantum Approximate Optimization Algorithm, which is an approximate version of the Quantum Adiabatic Algorithm (QAA)~\cite{farhi2000quantum}, leveraging the Adiabatic Theorem~\cite{Born1928} to find (approximate) solutions of unconstrained combinatorial optimization problems. For a binary objective function \( f: \{0, 1\}^n \rightarrow \mathbb{R} \), QAOA operates as follows~\cite{farhi2014quantum}:
\begin{enumerate}
    \item Map objective values to the energy levels of a Hamiltonian \( H_C = \sum_x f(x) \ket{x}\!\bra{x} \).
    \item Initialize the system in the ground state of a Hamiltonian, e.g., \( H_M = -\sum_{i=1}^n \sigma_i^x \) with ground state \( \ket{+}^{\otimes n} \).
    \item Simulate the time-ordered \( \exp(-i \int_0^T H_s(t) \, dt) \), where the continuous-time Hamiltonian \( H_s(t) = (1 - s(t)) H_M + s(t) H_C \) guides the adiabatic transition for \( s: [0, T] \rightarrow [0, 1] \) transitioning from 0 to 1.
    \item Measure the final state \( \ket{\psi} \) and map it back to a solution for \( f \).
\end{enumerate}
To implement this time evolution on gate-based quantum computers, the QAOA discretizes the continuous Hamiltonian into \( p \) segments \( H_s(1/T), \dots, H_s(T) \) and applies a first-order Suzuki-Trotter approximation. This yields the unitary operator:
\begin{equation}
    U(\beta, \gamma) = U_M(\beta_p) U_C(\gamma_p) \ldots U_M(\beta_1) U_C(\gamma_1),
\end{equation}
where \( \beta_i \) and \( \gamma_i \) are introduced as variables that control the evolution rate, and \( U_M(\beta_i) = e^{-i \beta_i H_M} \), \( U_C(\gamma_i) = e^{-i \gamma_i H_C} \). As \( p \rightarrow \infty \), \( U(\beta, \gamma) \) approximates the adiabatic evolution increasingly accurately. A plethora of efficient training procedures and initializations for the parameters has been established in literature, cf. Refs.~\cite{apte2025iterativeinterpolationschedulesquantum,sack2021}.

\subsection{QuSO solving}
We now introduce the overall setup of the proposed QuSO solver as defined in \Cref{thm:quso}.

\begin{theorem}[QuSO Solver Architecture~\protect{\cite[Thm.~10]{stein2024exponentialquantumspeedupsimulationbased}}]\label{thm:quso}
    The circuit displayed in \Cref{fig:quso} implements a quantum algorithm for solving a QuSO problem of the form $\argmin_x u(s(x))$ given an oracle $\textnormal{QSim}\ket{x}\ket{0}^{\otimes m}=\ket{x}\ket{u(s(x))}$.
\end{theorem}

\begin{figure*}[hbtp]
\centering
\input{tikz/quso}
\caption{Quantum circuit addressing a QuSO problem of the form \( \arg\min_x f(x) = u(s(x)) \), as outlined in \Cref{thm:quso}. A possible implementation of QSim is detailed in Ref.~\protect{\cite[Fig.~9]{stein2024exponentialquantumspeedupsimulationbased}}. The \(\begin{quantikz}\ground{}\end{quantikz}\) symbol indicates ancillary qubits to be reinitialized in the \( \ket{0} \) state, which can be reused for subsequent calculations. This circuit is taken from Ref.~\cite{stein2024exponentialquantumspeedupsimulationbased} without alteration.}
\label{fig:quso}
\end{figure*}

If the simulation component of the QuSO problem takes the form of a system of linear equations (as assumed for the rest of this paper and applicable to the problem of interest), exponential speedups are possible if the following conditions are met~\cite{stein2024exponentialquantumspeedupsimulationbased}:
\begin{enumerate}
    \item The SLE is sparse and well-conditioned.
    \item The dependency of the SLE on decision variables allows efficient input to a quantum linear system solver.
    \item Summary statistic information from the SLE solution can be extracted efficiently.
\end{enumerate}

For the sake of brevity, we refer the reader to Ref. \cite{stein2024exponentialquantumspeedupsimulationbased} for any further details on the simulation component QSim, as that section of the algorithm will not be part of a more thorough consideration within this manuscript, since we concentrate on classical simulations of the algorithm which can exploit precomputation of the diagonal cost unitary without any additional ancillary qubits needed (cf.~\Cref{subsec:implementation}). Notably, this is done only for evaluation purposes and does not hinder any future QPU implementations.

\section{Methodology}
\label{sec:methodology}
In this section, we define the specific version of the unit commitment problem addressed in this paper and propose a simplified classical circuit simulation using extensive precomputation combined with a diagonalized form of the QAOA for the sake of extensive benchmarking.

\subsection{The Unit Commitment Problem}
The unit commitment problem is a MINLP-type optimization problem, concerned with determining which power generators should be active within a power grid to meet a given estimated load at the lowest possible cost. The associated simulation problem, known as the power flow problem, involves calculating the power flow through each transmission line, \( \vec{\rho}_{ij} \), based on the power input and output at each node (referred to as buses) in the power grid, \( \vec{p}_i \). The costs of this power flow problem can be expressed as \( \langle \vec{c}_{ij}, |\vec{\rho}(x)_{ij}|\rangle  \), where $\vec{c}_{ij}$ represents the linear cost factor for transmitting power over each transmission line. Although the unit commitment problem includes many additional cost factors and constraints in practical settings (which can, in principle, be incorporated into the proposed QuSO solver, cf. Ref.~\cite{stein2024exponentialquantumspeedupsimulationbased}), this paper focuses on a simplified version to serve as a proof of concept. Specifically, we investigate a widely used linear approximation of AC power flow (cf. \Cref{thm:DCapprox}).

We begin by defining the general components and features of a power grid in \Cref{def:powergrid} and directly follow up with computation of the power flow over the transmission lines in \Cref{thm:DCapprox}.


\begin{definition}[Power Grid]\label{def:powergrid}
    We define a power grid through the 5-tuple $(G,L,\vec{b}_{ij}, \vec{p}_i,\vec{c}_{ij})$, where $G$ denotes the set of generators and $L$ the set of loads. Together, these form the set of buses in the grid, $V\coloneqq G\cup L$. Each node $i$ is assigned a power value $p_i$, which is positive for generators and negative for loads, and corresponds to their power in-/output to the grid. The transition lines between the buses are modeled as edges $E\subseteq V\times V$ of the graph
    $\mathcal{G}=(V,E)$. Each transition line $(i,j)\in E$ has a corresponding susceptance $b_{ij}$ and a linear base factor of costs to send power over this line $c_{ij}$.
\end{definition}


\begin{theorem}[DC Power Flow Approximation, informal, ~\protect{\cite{wood2013power}}]\label{thm:DCapprox}
    Assuming that (i) the total load is low, (ii) the resistances in the transmission lines are negligible, (iii) the voltage magnitudes are close to nominal, and (iv) the difference between the voltage angles is small for a given AC power grid $(G,L,\vec{b}_{ij}, \vec{p}_i,\vec{c}_{ij})$, we can compute the power flowing over all transmission lines $\rho_{ij}$ by solving the system of linear equations
    \begin{equation}
        p_i = \sum_j b_{ij} (\theta_i-\theta_j) \quad \forall i\in V
    \end{equation}
    for the so-called voltage angles $\theta_i$, and then inserting them into the straightforward expression
    \begin{equation}
        \rho_{ij}=b_{ij} (\theta_i-\theta_j).
    \end{equation}
\end{theorem}

This already allows us to formally define the specific unit commitment problem focused on the power flow costs in \Cref{def:UCP}.

\begin{definition}[Unit Commitment Problem (UCP)]\label{def:UCP}
     Given a power grid $(G,L,\vec{b}_{ij}, \vec{p}_i,\vec{c}_{ij})$, we define the problem of unit commitment as the optimization that asks which generators $G(x)\coloneqq \left\lbrace i \in G : x_i = 1\right\rbrace$ to operate in order to minimize the overall power transmission costs using the decision variables $x\in\left\lbrace 0,1\right\rbrace^{|G|}$, while the net power input is roughly equal to the net power output, i.e.,
    \begin{equation}
        \argmin_{x\in\left\lbrace 0,1\right\rbrace^{|G|}} \langle \vec{c}_{ij}, |\vec{\rho}(x)_{ij}|\rangle \textnormal{ s.t. } \sum_i \vec{p}(x)_i \approx 0,
    \end{equation}\label{eq:opt_problem}where $\vec{p}(x)_i \coloneqq x_i\vec{p}_i $ if $ i \in G $, and $\vec{p}(x)_i \coloneqq \vec{p}_i $, otherwise; and $\vec{\rho}(x)_{ij}$ denotes the power flowing over all transmission lines when substituting $\vec{p}(x)_i$ for $\vec{p}_i$ in \Cref{thm:DCapprox}. 
    
\end{definition}

As shown in Ref.~\cite{stein2024exponentialquantumspeedupsimulationbased}, the associated cost function can be efficiently computed within the QuSO $\textnormal{QSim}$ framework, as formally restated in the following theorem.

\begin{theorem}[QSim implementation for the UCP~\protect{\cite[Thm.~13]{stein2024exponentialquantumspeedupsimulationbased}}]
    If $s(x)$ takes the form of a linear system of equations and $u(x)$ takes the form of an inner product as in \Cref{def:UCP}, we can implement the corresponding oracle $\textnormal{QSim}\ket{x}\ket{0}^{\otimes m}=\ket{x}\ket{u(s(x))}$ within a quantum circuit of depth $\tilde{\mathcal{O}}(\polylog(N)\kappa^{*}s/\varepsilon\delta)$ using $\tilde{\mathcal{O}}(N^2)$ ancillary qubits, where $\varepsilon$ denotes the accuracy of extracting the summary statistic solution, and $0<\delta<1$ the success probability of the algorithm.
\end{theorem}

\subsection{Quantum Simulation Methodology}\label{subsec:implementation}
As the circuit proposed in \Cref{thm:quso} has components that inherently rely heavily on fault-tolerant quantum computing (e.g., the QSVT), all evaluations must still be performed using classical circuit simulators at the moment. This allows for substantial simplifications of the required subroutines within the algorithm, as its most complex component -- the cost unitary -- can be precomputed classically, resulting in a diagonal matrix with entries $\exp(-i \gamma_i u(s(x)))$ for all $x$. While this requires each possible solution to be computed in a brute force manner, hence increasing the computational complexity of the optimization algorithm exponentially, this step is crucial in reducing the number of ancillary qubits to zero, i.e., this approach only requires as many qubits as there are decision variables. As classical quantum circuit simulation is mostly bottlenecked with the number of available qubits, this allows us to test for the largest scale problems possible, which best indicate the eventual scaling of the proposed QuSO solver.

The other components of the proposed algorithm (cf. \Cref{fig:quso}) consist only of the state preparation routine and the mixer unitary. As we use the standard $X$-Mixer, together with the $\ket{+}^{\otimes n}$ state as its corresponding ground state, we simply have $U_S = H^{\otimes n}$ and $U_M(\beta_i) = e^{-i \beta_i X}$. The final step of cost computation displayed in \Cref{fig:quso} is not simulated inside the quantum circuit, as it is unnecessary in our implementation, based on the fact that we already have precomputed each cost value for all solutions.

\section{Experimental Setup}
\label{sec:experimental_setup}
In this section, we introduce the hyperparameters for the classical baseline and our quantum approach, the dataset and the evaluation metrics.

\subsection{Quantum Circuit Implementation and Parameter Training}
For the concrete implementation, we utilize PennyLane to build the QAOA circuit layer by layer. Due to the brute force computation of the diagonal cost Hamiltonian, rather than via operator strings, we are restriced to use the \texttt{default.qubit} device with density matrices/arrays.

In order for the QAOA algorithm to perform well, the respective parameters $(\gamma_i, \beta_i)$ have to be trained first. Based on a preliminary study, we selected the \texttt{AdamOptimizer} for this task. 
Rather than directly optimizing all $2p$ parameters, we adopt the Iterative Interpolation (II) schedule proposed by Apte \textit{et al.}~\cite{apte2025iterativeinterpolationschedulesquantum} to allow for efficient scaling in $p$. The core observation is that well-optimized QAOA angle schedules vary smoothly as a function of the normalized layer index $t = i / p \in [0, 1]$, meaning that they can be well approximated using only a small number of basis function coefficients $C$. We choose the basis functions to be Chebyshev polynomials $T_j$, which formally reads as
\begin{align}
    \gamma_i &= \sum_{j=1}^C u_j T_j (i/p), \textnormal{and} \\
    \beta_i &= \sum_{j=1}^C v_j T_j(i/p).
\end{align}
This reduces the effective optimization dimensionality from $2p$ to $2C$, where $C$ can be chosen such that $C \ll p$, potentially even remaining completely constant. Starting from a circuit of depth $p_0$, the algorithm iteratively increases depth in increments of $\Delta p$, re-expressing the current angles in the given basis and only optimizing the leading $C$ coefficients. When performance improvement saturates, which is measured by a threshold $\varepsilon$ not being exceeded for $\tau$ consecutive iterations, the number of coefficients $C$ is increased to capture finer features. The optimized coefficients are then interpolated by evaluating the series expansion at the new points $t_i = i/(p + \Delta p)$.

\subsection{Classical Baseline}
As a classical baseline we use Simulated Annealing (SA)~\cite{kirkpatrick1983optimization}, as it is an iterative, gradient-free search  algorithm that allows solving the simulation problem with specialized solvers like the conjugate gradient method as opposed to a MILP solver like CPLEX \cite{cplex} or Gurobi \cite{gurobi}, for which the number of equality constraints becomes excessively large for high dimensional simulation problems. 

The here employed version of Simulated Annealing begins with a random initial solution $x \in \mathcal{X}$ and conducts a local search. During this, it iteratively chooses a neighboring solution $x' \in \mathcal{N}(x)$ at random in each step, where $\mathcal{N}(x)$ is the set of candidates with hamming distance to the current candidate equal to 1. If the new candidate $x'$ improves the objective function, it is accepted unconditionally (i.e., $x \leftarrow x'$). Otherwise, if $x'$ has a worse objective value, it is accepted with probability $P(x,x',T) \coloneqq e^{-\left|C(x) - C(x')\right| / kT}$. Here, $k$ is the Boltzmann constant, and $T_i$ represents the temperature at iteration $i$, which is updated according to a predefined annealing schedule. For the sake of simplicity and in order to use the temperature as a direct control parameter, we set $k = 1$. Furthermore, we employ the commonly used geometric annealing schedule, where the temperature decreases exponentially according to $T_i \leftarrow \alpha T_{i-1}$, with initial temperature $T_0 = 1000$ and final temperature $T_{f} = 1$, with decay parameter $\alpha \in (0,1)$~\cite{Yaghout_Nourani_1998}. For our experiments, we explore a range of temperature iteration steps starting at $10$ and increasing to $512$ via geometric progression. The algorithm returns the best solution identified over a specified maximum number of iterations and is repeated for a specific number of times to ensure statistical relevance (for our experiments, we use 1000 samples). As the cost function is already precomputed for every solution candidate in the course of our QAOA implementation, a mere look-up can be used during each temperature iteration for evaluation purposes.



\subsection{Dataset}
For the desired proof-of-concept evaluation, we sample random connected graphs that serve as simplified models of power grids. Each vertex is assigned a random load power. For a problem with $N$ qubits, we randomly designate $N$ vertices as generators, each providing power generation. Both load and generation values are drawn from a uniform distribution, $\mathcal{U}(0,1)$. 
Each edge is assigned a reactances $X$ and resistances $R$, also sampled uniformly, subject to the constraint $R \ll X$. This ensures that the network satisfies the assumptions of the DC power flow approximation (cf. \Cref{thm:DCapprox}).

\subsection{Evaluation Metrics}
The main metric used in this paper is Random Adjusted Approximation Ratio (RAAR) \cite{Bucher_2025}. As an extension of Approximation Ratio \cite{bucher2024robustbenchmarkingquantumoptimization}, it quantifies how close the best found solution is to an optimal solution, yet additionally accounts for random guessing. It is formularized as follows:
\begin{equation}
    \text{RAAR} =  \dfrac{\langle f(x) \rangle - \langle \psi \vert f(x) \vert \psi \rangle}{\langle f(x) \rangle -\min_xf(x)},
    \label{eq:RAAR}
\end{equation}
where $\langle f(x) \rangle$ corresponds to the random average of the cost $f(x)$. In practice, $\langle f(x) \rangle$ reduces to the mean cost over all solution candidates, while the expectation $\langle \psi \vert f(x) \vert \psi \rangle$ is extracted from summary statistics in the quantum case and is defined as the mean cost of the best solutions found over multiple (here, 1000) runs in the classical case. 
$\text{RAAR} = 1$ is the upper bound of the metric, which is reached iff. $\langle \psi \vert f(x) \vert \psi \rangle = \text{min}_x f(x)$, while $\text{RAAR} = 0$ corresponds to the expected solution quality of random sampling.

The other employed main metric is Time to Solution (TTS), which we define as the total number of computing steps, proportional to overall runtime, required to yield an optimal solution with 99\% certainty, given that the probability of sampling an optimal solution is $P_s$ (cf. Ref.~\cite{bucher2024robustbenchmarkingquantumoptimization}):
\begin{equation}
    \text{TTS}(s) = \textbf{L}(s)\cdot \left\lceil\frac{\log(0.01)}{\log(1.0 - P_s)}\right\rceil,
\label{eq:tts}
\end{equation}
where $\textbf{L}(s)$ is a measure proportional to the depth of the algorithm and thus to its execution time.
In accordance with this, we further define TTS$^*$ in Eq. \eqref{eq:tts_star} as the minimum TTS with respect to $s$, i.e., the TTS in the optimal configuration of the respective algorithm:

\begin{equation}
    \text{TTS}^* = \min_s \\ \text{TTS($s$)}.
\label{eq:tts_star}
\end{equation}

We define $\textbf{L}(s)$ as the number of layers for the QAOA and the number of temperature iterations for Simulated Annealing correspondingly. For a better end-to-end comparison, the resulting TTS$^*$ is later multiplied by the respective complexity of solving the underlying simulation problem, i.e., $\mathcal{O}(\text{polylog}(N)\kappa^* s / (\epsilon \delta))$ in the quantum case and $\mathcal{O}(N s \kappa \log(1/\epsilon))$ in the classical case. Details on the exact values used for the comparison are provided in \Cref{subsec:TTS}. 

\section{Results}
\label{sec:results}

\begin{figure*}[t]
    \centering

    \subfloat[QAOA performance for initial circuit depths $p_0$, averaging over all loads]{\includegraphics[width=0.48\linewidth]{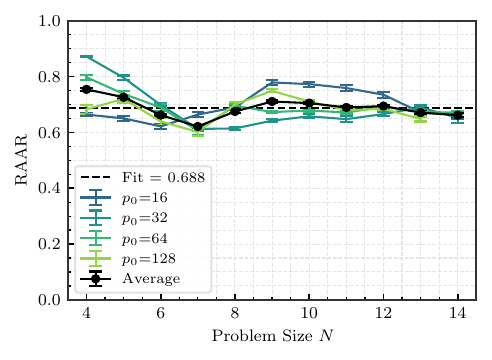}\label{fig:raar_qaoa_N}}
    \hfill
    \subfloat[SA performance for temperature iterations $p$, averaged over all loads]{\includegraphics[width=0.48\linewidth]{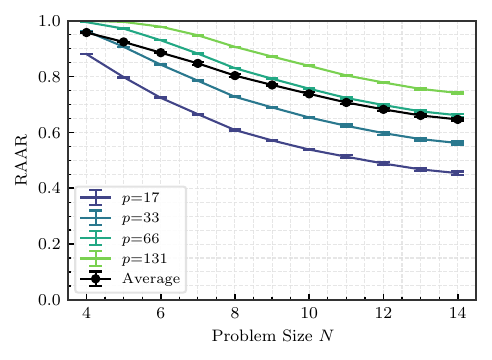}\label{fig:raar_sa_N}}

    \vspace{1em}


    \caption{Solution quality of QAOA (left) and Simulated Annealing (right) measured as RAAR vs. problem size $N$}
    \label{fig:raar}
\end{figure*}

This section is structured in three parts, displaying results on the solution quality, the runtime complexity, a performance comparison on the different problem instances in terms of the approximation ratios, and finally the time to solution.

\subsection{Evaluation Details}\label{subsec:evaldetails}
For each combination of load factor and problem size, we generated 30 random power grid instances. In the quantum case, one instance per (problem size, load) pair was selected at random to train the QAOA angles via the Iterative Interpolation (II) schedule. The resulting angles were then used for all 30 instances, without retraining. This strategy is central to the computational appeal of the approach, since training QAOA angles for every single instance is computationally expensive. The diagonal cost Hamiltonian $H_C$ is normalized to $[0, 1]$ via min-max scaling. We consider problem sizes $4 \leq N \leq 14$ and initial circuit depths $p_0 \in \{ 16, 32, 64, 128 \}$, with the II schedule extending up to a maximum depth of $p_\text{max} = 2 p_0$ in increments of $\Delta p = p_0/8$, using only $C = p_0 / 4$ Chebyshev coefficients. The initial angles $(\gamma_i, \beta_i)$ are given by a linear ramp in line with Ref.~\cite{sack2021}. Parameters are optimized using the Adam optimizer for 25 steps per interpolation stage, with stagnation detected after $\tau = 3$ consecutive iterations below the improvement threshold $\varepsilon = 1$. In line with Ref.~\cite{apte2025iterativeinterpolationschedulesquantum}, training terminates early once a target approximation ratio of $\text{AR} = 80\%$ is reached. Note that this latter training target is distinct from the $\text{RAAR}$ metric used for evaluation.

As results are averaged across structurally diverse power grid instances, raw variance tends to be large. We therefore use the standard error of the mean rather than the standard deviation, when drawing error bars. The standard error better reflects the uncertainty in the estimated average performance rather than the spread across instances.

\subsection{Solution Quality}
\label{section:solution_quality}
The solution quality of the QAOA depends on both problem size $N$ and circuit depth $p$. In \Cref{fig:raar}, we evaluate performance across varying $N$ and initial depths $p_0$, applying the trained angles to all 30 instances and averaging over both instances and all 10 load configurations to obtain \Cref{fig:raar_qaoa_N}.
No clear ordering emerges among the different $p_0$ values -- all configurations fluctuate within the $60\%-90\%$ RAAR regime, which is unsurprising given the averaging over structurally diverse instances and load configurations. Notably, solution quality remains largely stable with increasing problem size $N$. This leads to a compelling observation: since only $C = p_0 / 4$ Chebyshev coefficients are optimized, as few as $C=4$ coefficients are sufficient to achieve competitive performance across all problem sizes up to $N = 14$.
Whether or, more likely, \emph{when} solution quality eventually degrades for larger $N$ remains an open question. On the other hand, it would be very interesting to explore how our approach would have to be changed in order to benefit from deeper QAOA layers. One compelling possibility is the termination threshold for the training, that is set to 80\% AR (cf. \Cref{subsec:evaldetails}), which should be investigated more closely in future work.  
Averaging additionally over all $p_0$ values yields the drawn black data points, where a constant fit gives an overall average RAAR of $69\%$.
The absence of a clear $p$-dependence is further confirmed in \Cref{fig:qoa_raar_2}, where RAAR is plotted against circuit depth $p$ for selected load configurations, averaged over all problem sizes $N$. Note that due to the interpolation step, the circuit depth after one iteration is given by $p = p_0 + \Delta p$.
 Unlike the quantum algorithm, Simulated Annealing clearly experiences a monotonic improvement in solution quality dependent on the number of temperature iterations, as can be observed both in \Cref{fig:raar_sa_N} and \Cref{fig:qaoa_raar_2}. The overall relation between temperature iterations and RAAR appears to be of logarithmic nature, providing a significant increase in performance from 1 to 100, while slowly approaching near-optimal solution quality at 512 for all load factors. Please note, that the $p$-values in \Cref{fig:raar_sa_N} do not fully match the $p_0$-values in \Cref{fig:raar_qaoa_N}, as the geometric progression employed for the temperature iterations does not yield powers of 2, and thus the closest available data points were used. Furthermore, \Cref{fig:raar_sa_N} demonstrates a graceful degradation of RAAR with increasing problem size. The observed trend corresponds to an exponential decay, which, while for relatively large numbers of temperature iterations ($p=66$, $p=131$), still allows us to obtain acceptable solutions (depending on requirements) with a RAAR of over 70\%, the performance for smaller numbers ($p=17$) decays more severely. In direct comparison, the QAOA reports stable results across all problem instances, even for small number of initial layers (e.g. $p_0=16$), whereas the opposite is the case for SA. Smaller number of temperature iterations (e.g. $p=17$) experience a non-negligible decrease in solution quality with increasing problem size. On the other hand, increasing the number of temperature iterations has a direct impact on RAAR, while a mere doubling of QAOA layers does not appear to achieve a better performance. Here, further or different parameter optimization might be a possible solution.

\begin{figure*}[t]
    \centering

    \subfloat[QAOA performance for selected loads, averaging over all $N$.]{\includegraphics[width=0.48\linewidth]{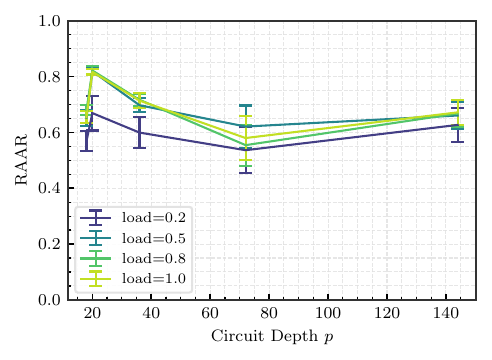}\label{fig:qoa_raar_2}}
    \hfill
    \subfloat[SA performance averaged over all problem sizes $N$.]{\includegraphics[width=0.48\linewidth]{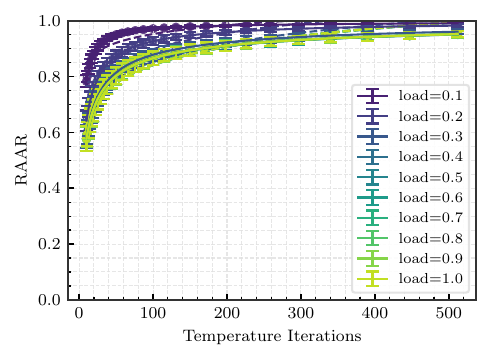}\label{fig:qaoa_raar_2}}
    
    \caption{Solution quality of QAOA (left) and Simulated Annealing (right) as a function of computational effort.}
    \label{fig:raar2}
\end{figure*}

\subsection{Time to Solution}\label{subsec:TTS}
We now analyze the minimum Time to Solution TTS$^*$ to investigate whether the QAOA-based approach offers an end-to-end computational runtime advantage over the classical baseline. 
For the purpose of this comparison, we adopt the following assumptions on the coefficients determining the complexity of the simulation component: we set the maximally allowed error $\epsilon$ to $0.001$, and assume $\kappa = \kappa^*$, such that the condition number cancels between the two expressions (which is reasonable, as $\kappa^*$ and $\kappa$ are only separated by constants in practice~\cite{stein2024exponentialquantumspeedupsimulationbased}). Further, we also assume $\delta$ to be constant (e.g., $0.5$), which fully suffices for BQP algorithms in practice\cite{Nielsen_Chuang_2010}. Finally, we assume $\polylog$ to be quadratic, as the true polynomial is likely of low degree (cf. Ref.~\cite{stein2024exponentialquantumspeedupsimulationbased}).
The respective results are shown in \Cref{fig:tts} for selected load configurations, averaged over all instances and circuit depths $p$. Exponential fits in the log-scale plot reveal that lower load configurations exhibit steeper scaling, indicating that they constitute harder problem instances. Averaging over all 10 load configurations yields \Cref{fig:qaoa_tts_avg}, from which we determine the overall scaling behavior of the QAOA to be $2.06^N$.
Similarly to the observations regarding the solution quality in \Cref{section:solution_quality}, the employed classical algorithm shows a uniform performance across all load factors (\Cref{fig:sa_tts}), indicating an invariance of Simulated Annealing to the specific problem, assuming the complexity of the candidate space, i.e. the problem size, remains constant. The scaling behavior of SA minimally exceeds a basis of 2 for all load configurations, yielding an average of $2.032^N$, as can be observed in \Cref{fig:sa_tts_avg}. Thus, on average, QAOA was not able to outperform the classical baseline regarding the Time to Solution. However, most interestingly, QAOA's performance incorporates a much higher variance across problem classes. Selected classes, corresponding to the higher end of load factors, yield a significantly slower scaling behavior. For a load factor of 1 the effective speed-up compared to Simulated Annealing was approximately $\frac{2.055^N}{1.322^N} \approx 1.55^N$. On the other hand, in compliance with the No-Free-Lunch theorem, the effective slow-down for a load factor of 0.2 was approximately $\frac{2.023^N}{2.524^N} \approx 0.8^N$.

An aspect one needs to keep in mind regarding these results, however, is that our analysis was confined to $\mathsf{NP}$-hard problems. Therefore, it is unsurprising that both algorithms scale exponentially, and most importantly, that the sub-exponential simulation complexity factors $\log^2(N)\epsilon^{-1}$ and $N\log(\epsilon^{-1})$ merely influence the overall scaling behavior slightly. The impact of the quantum speedup for the simulation component would certainly be much larger for problem instances that allow for solving the optimization in polynomial time.


\begin{figure*}[t]
    \centering

    \subfloat[QAOA performance for selected loads, averaged over all layers $p$. Dashed line indicates exponential fit.]{\includegraphics[width=0.48\linewidth]{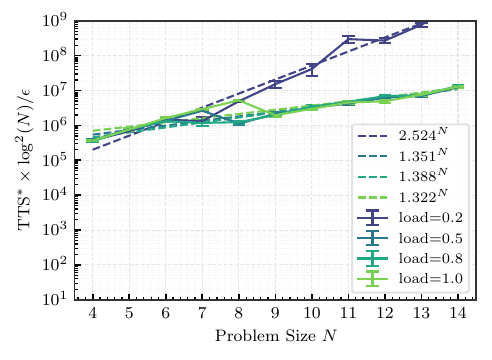}\label{fig:qaoa_tts}}
    \hfill
    \subfloat[SA performance for selected loads, averaged over all temperature iterations, on a logarithmic scale. Dashed lines correspond to exponential fits.]{\includegraphics[width=0.48\linewidth]{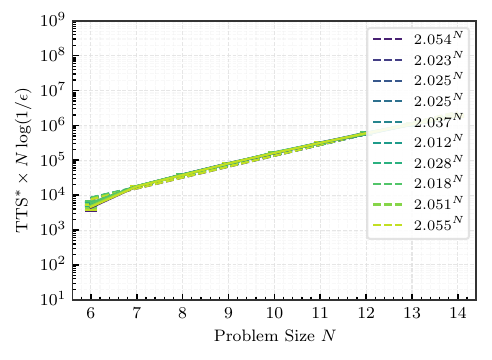}\label{fig:sa_tts}}

    \vspace{1em}

    \subfloat[Corresponding average across all 10 loads.]{\includegraphics[width=0.48\linewidth]{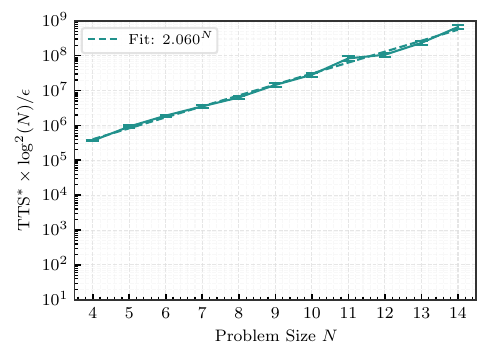}\label{fig:qaoa_tts_avg}}
    \hfill
    \subfloat[SA performance averaged over all temperature iterations and loads, on a logarithmic scale. The dashed line corresponds to an exponential fit.]{\includegraphics[width=0.48\linewidth]{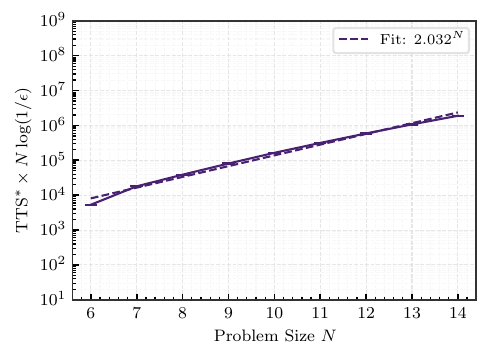}\label{fig:sa_tts_avg}}

    \caption{Complexity-normalized $\text{TTS}^*$ with $\text{TTS}^* \times \log^2 (N) /\epsilon$ for QAOA (left) and $\text{TTS}^* \times N \log (1/\epsilon)$ for Simulated Annealing (right).}
    \label{fig:tts}
\end{figure*}

\section{Discussion}
\label{sec:discussion}
Although in our analysis the QAOA was not able to achieve near-optimal approximation ratios, unlike Simulated Annealing in the lower end of problem sizes, we conclude that this only stems from insufficient and/or unsuitable parameter optimization. Furthermore, the quantum approach, while showing comparable performance regarding time to solution, achieved significant sub-exponential speed-ups for certain load factors of the unit commitment problem. This is a most interesting finding that showcases the potential of quantum-based simulation approaches, even if for other load factors, the QAOA experienced slight slow-downs. As the quantum speed-up for the simulation itself vanishes under the dominance of the exponential scaling of the algorithm for NP-hard problems, we conclude that the observed speed-ups would most likely stand out further for polynomial instances. 
Overall, it has been shown that the QAOA can adopt desirable properties for simulation-based optimization problems, most importantly, significant speed-ups compared to a classical baseline for specific subclasses of problem instances (here: heavy loads on the power grid).

\section{Conclusion}
\label{sec:conclusion}
Our numerical experiments demonstrated that quantum simulation-based optimization approaches can offer an end-to-end runtime advantage over their classical counterparts. We find that QAOA is capable of achieving noteworthy improvements in computational complexity for certain problem instances while keeping solution quality stable across various problem sizes, even for circuit layer depths as small as 16. Due to our extensive benchmarking across various problem sizes, corresponding random instances and load factors, we further expect our insights to generalize to other instances of the unit commitment problem.
A noteworthy detail from our analysis is that the QAOA was not able to reach near-optimal solution quality in most cases. This issue likely stems from the Iterative Interpolation algorithm employed for optimizing the variational parameters. While it offers a straightforward way to optimize variational parameters for deep QAOA circuits, it lacked noticeable improvements for deeper circuits -- likely due to improper hyperparameter optimization. 
Future work should definitely conduct a broader hyperparameter search and expand the evaluation to larger problem instances.
Given the immense range of industrially and academically relevant QuSO problems, our results underscore their potential for practically relevant quantum advantage, as we were able to expand the existing results on speedups purely for the simulation component to an end-to-end speedup that includes the outer optimization component.

\section*{Acknowledgment}
This work has been supported by the LMU Sustainability Fund (EfOiE), the BMFTR (QuCUN, QuaRDS, CAQAO), the Munich Quantum Valley (K5, K7), and the Bavarian StMWi (6GQT). The sole responsibility for the report’s contents lies with the authors.

\section*{Appendix}
\label{sec:appendix}

The code can be accessed via this Github link: \url{https://github.com/DerEinverleiber/UCP-PaperV2}

\bibliographystyle{IEEEtranDoi}  
\bibliography{bstcontrol,references} 

@IEEEtranBSTCTL{BSTcontrol,
  CTLuse_forced_etal       = "no",
  CTLmax_names_forced_etal = "10",
  CTLnames_show_etal       = "10",
  CTLuse_url               = "yes",
  CTLdash_repeated_names   = "no",
  CTLname_url_prefix       = ""
}

@book{Nielsen_Chuang_2010,
    place={Cambridge},
    title={Quantum Computation and Quantum Information: 10th Anniversary Edition},
    publisher={Cambridge University Press},
    author={Nielsen, Michael A. and Chuang, Isaac L.},
    year={2010},
    doi={10.1017/CBO9780511976667}
}

@misc{farhi2000quantum,
    title={Quantum Computation by Adiabatic Evolution}, 
    author={Edward Farhi and Jeffrey Goldstone and Sam Gutmann and Michael Sipser},
    year={2000},
    eprint={quant-ph/0001106},
    archivePrefix={arXiv},
    primaryClass={quant-ph},
}

@misc{farhi2014quantum,
    title={A Quantum Approximate Optimization Algorithm}, 
    author={Edward Farhi and Jeffrey Goldstone and Sam Gutmann},
    year={2014},
    eprint={1411.4028},
    archivePrefix={arXiv},
    primaryClass={quant-ph}
}

@article{Born1928,
    author = {Born, M and Fock, V},
    doi = {10.1007/BF01343193},
    issn = {0044-3328},
    journal = {Zeitschrift f{\"{u}}r Phys.},
    number = {3},
    pages = {165--180},
    title = {{Beweis des Adiabatensatzes}},
    url = {https://doi.org/10.1007/BF01343193},
    volume = {51},
    year = {1928}
}

@article{sack2021,
	title = {Quantum annealing initialization of the quantum approximate optimization algorithm},
	volume = {5},
	doi = {10.22331/q-2021-07-01-491},
	journal = {Quantum},
	author = {Sack, Stefan H. and Serbyn, Maksym},
	month = jul,
	year = {2021},
	pages = {491}
}

@misc{gilyen2018quantum,
    title={Quantum singular value transformation and beyond: exponential improvements for quantum matrix arithmetics},
    author={Gily{\'e}n, Andr{\'a}s and Su, Yuan and Low, Guang Hao and Wiebe, Nathan},
    eprint={1806.01838},
    year={2018},
    archivePrefix={arXiv},
    primaryClass={quant-ph}
}

@article{HHL,
  title = {Quantum Algorithm for Linear Systems of Equations},
  author = {Harrow, Aram W. and Hassidim, Avinatan and Lloyd, Seth},
  journal = {Phys. Rev. Lett.},
  volume = {103},
  issue = {15},
  pages = {150502},
  numpages = {4},
  year = {2009},
  month = {Oct},
  publisher = {American Physical Society},
  doi = {10.1103/PhysRevLett.103.150502},
}

@InProceedings{Ambainis12,
  author =	{Ambainis, Andris},
  title =	{{Variable time amplitude amplification and quantum algorithms for linear algebra problems}},
  booktitle =	{29th International Symposium on Theoretical Aspects of Computer Science (STACS 2012)},
  pages =	{636--647},
  series =	{Leibniz International Proceedings in Informatics (LIPIcs)},
  ISBN =	{978-3-939897-35-4},
  ISSN =	{1868-8969},
  year =	{2012},
  volume =	{14},
  editor =	{D\"{u}rr, Christoph and Wilke, Thomas},
  publisher =	{Schloss Dagstuhl -- Leibniz-Zentrum f{\"u}r Informatik},
  address =	{Dagstuhl, Germany},
  doi =		{10.4230/LIPIcs.STACS.2012.636},
}

@INPROCEEDINGS{Berry15,
  author={Berry, Dominic W. and Childs, Andrew M. and Kothari, Robin},
  booktitle={2015 IEEE 56th Annual Symposium on Foundations of Computer Science}, 
  title={Hamiltonian Simulation with Nearly Optimal Dependence on all Parameters}, 
  year={2015},
  volume={},
  number={},
  pages={792-809},
  doi={10.1109/FOCS.2015.54}}

@article{Childs17,
author = {Childs, Andrew M. and Kothari, Robin and Somma, Rolando D.},
title = {Quantum Algorithm for Systems of Linear Equations with Exponentially Improved Dependence on Precision},
journal = {SIAM Journal on Computing},
volume = {46},
number = {6},
pages = {1920-1950},
year = {2017},
doi = {10.1137/16M1087072},
}

@article{SSO19,
  title = {Quantum Algorithms for Systems of Linear Equations Inspired by Adiabatic Quantum Computing},
  author = {Suba{\c{s}}{\i}, Y. and Somma, Rolando D. and Orsucci, Davide},
  journal = {Phys. Rev. Lett.},
  volume = {122},
  issue = {6},
  pages = {060504},
  numpages = {5},
  year = {2019},
  month = {Feb},
  publisher = {American Physical Society},
  doi = {10.1103/PhysRevLett.122.060504},
}

@article{Lin2020optimalpolynomial,
  doi = {10.22331/q-2020-11-11-361},
  url = {https://doi.org/10.22331/q-2020-11-11-361},
  title = {Optimal polynomial based quantum eigenstate filtering with application to solving quantum linear systems},
  author = {Lin, Lin and Tong, Yu},
  journal = {{Quantum}},
  issn = {2521-327X},
  publisher = {{Verein zur F{\"{o}}rderung des Open Access Publizierens in den Quantenwissenschaften}},
  volume = {4},
  pages = {361},
  month = nov,
  year = {2020}
}

@article{posDefQLSP,
  doi = {10.22331/q-2021-11-08-573},
  title = {On solving classes of positive-definite quantum linear systems with quadratically improved runtime in the condition number},
  author = {Orsucci, Davide and Dunjko, Vedran},
  journal = {{Quantum}},
  issn = {2521-327X},
  publisher = {{Verein zur F{\"{o}}rderung des Open Access Publizierens in den Quantenwissenschaften}},
  volume = {5},
  pages = {573},
  month = nov,
  year = {2021}
}

@article{balabanov1996topology,
  title={Topology optimization of transport wing internal structure},
  author={Balabanov, Vladimir O and Haftka, Raphael T},
  journal={Journal of aircraft},
  volume={33},
  number={1},
  pages={232--233},
  year={1996},
  doi = {10.2514/6.1994-4414},
}

@book{myers2016response,
  title={Response surface methodology: process and product optimization using designed experiments},
  author={Myers, Raymond H and Montgomery, Douglas C and Anderson-Cook, Christine M},
  year={2016},
  publisher={John Wiley \& Sons},
  url={https://www.wiley.com/en-us/Response+Surface+Methodology%3A+Process+and+Product+Optimization+Using+Designed+Experiments%2C+4th+Edition-p-9781118916018}
}

@article{Gaitan2020,
author = {Gaitan, Frank},
doi = {10.1038/s41534-020-00291-0},
issn = {2056-6387},
journal = {npj Quantum Inf.},
number = {1},
pages = {61},
title = {{Finding flows of a Navier–Stokes fluid through quantum computing}},
url = {https://doi.org/10.1038/s41534-020-00291-0},
volume = {6},
year = {2020}
}

@misc{stein2024exponentialquantumspeedupsimulationbased,
      title={Exponential Quantum Speedup for Simulation-Based Optimization Applications}, 
      author={Jonas Stein and Lukas Müller and Leonhard Hölscher and Georgios Chnitidis and Jezer Jojo and Afrah Farea and Mustafa Serdar Çelebi and David Bucher and Jonathan Wulf and David Fischer and Philipp Altmann and Claudia Linnhoff-Popien and Sebastian Feld},
      year={2024},
      eprint={2305.08482},
      archivePrefix={arXiv},
      primaryClass={quant-ph},
      url={https://arxiv.org/abs/2305.08482},
      doi={10.48550/arXiv.2305.08482
}
}

@misc{montanaro2024quantumspeedupssolvingnearsymmetric,
      title={Quantum speedups in solving near-symmetric optimization problems by low-depth QAOA}, 
      author={Ashley Montanaro and Leo Zhou},
      year={2024},
      eprint={2411.04979},
      archivePrefix={arXiv},
      primaryClass={quant-ph},
      url={https://arxiv.org/abs/2411.04979}, 
      doi={10.48550/arXiv.2411.04979}
}

@inproceedings{10.1145/268437.268460,
author = {Carson, Yolanda and Maria, Anu},
title = {Simulation optimization: methods and applications},
year = {1997},
isbn = {078034278X},
publisher = {IEEE Computer Society},
address = {USA},
url = {https://doi.org/10.1145/268437.268460},
doi = {10.1145/268437.268460},
booktitle = {Proceedings of the 29th Conference on Winter Simulation},
pages = {118–126},
numpages = {9},
location = {Atlanta, Georgia, USA},
series = {WSC '97}
}

@article{kirkpatrick1983optimization,
	title = {Optimization by {Simulated} {Annealing}},
	volume = {220},
	url = {https://www.science.org/doi/10.1126/science.220.4598.671},
	linkdoi = {10.1126/science.220.4598.671},
	number = {4598},
	urldate = {2024-04-21},
	journal = {Science},
	author = {Kirkpatrick, S. and Gelatt, C. D. and Vecchi, M. P.},
	month = may,
	year = {1983},
	note = {Publisher: American Association for the Advancement of Science},
	pages = {671--680},
}

@article{Yaghout_Nourani_1998,
	title = {A comparison of simulated annealing cooling strategies},
	volume = {31},
	issn = {0305-4470, 1361-6447},
	url = {https://iopscience.iop.org/article/10.1088/0305-4470/31/41/011},
	linkdoi = {10.1088/0305-4470/31/41/011},
	number = {41},
	urldate = {2024-04-21},
	journal = {Journal of Physics A: Mathematical and General},
	author = {Nourani, Yaghout and Andresen, Bjarne},
	month = oct,
	year = {1998},
	pages = {8373--8385},
}

@INPROCEEDINGS{10821080,
  author={Stein, Jonas and Blenninger, Jonas and Bucher, David and Eder, Peter Josef and Çetiner, Elif and Zorn, Maximilian and Linnhoff-Popien, Claudia},
  booktitle={2024 IEEE International Conference on Quantum Computing and Engineering (QCE)}, 
  title={CUAOA: A Novel CUDA-Accelerated Simulation Framework for the QAOA}, 
  year={2024},
  volume={2},
  number={},
  pages={280-285},
  doi={10.1109/QCE60285.2024.10292},
eprint={2407.13012},
  archivePrefix={arXiv},}

@article{hestenes1952methods,
  title={Methods of conjugate gradients for solving linear systems},
  author={Hestenes, Magnus R and Stiefel, Eduard and others},
  journal={Journal of research of the National Bureau of Standards},
  volume={49},
  number={6},
  pages={409--436},
  year={1952}
}

@misc{bucher2024robustbenchmarkingquantumoptimization,
      title={Towards Robust Benchmarking of Quantum Optimization Algorithms}, 
      author={David Bucher and Nico Kraus and Jonas Blenninger and Michael Lachner and Jonas Stein and Claudia Linnhoff-Popien},
      year={2024},
      eprint={2405.07624},
      archivePrefix={arXiv},
      primaryClass={quant-ph},
      url={https://arxiv.org/abs/2405.07624}, 
}

@article{summary_statistic,
  title = {Quantum Algorithm for Linear Systems of Equations},
  author = {Harrow, Aram W. and Hassidim, Avinatan and Lloyd, Seth},
  journal = {Phys. Rev. Lett.},
  volume = {103},
  issue = {15},
  pages = {150502},
  numpages = {4},
  year = {2009},
  month = {Oct},
  publisher = {American Physical Society},
  doi = {10.1103/PhysRevLett.103.150502},
  url = {https://link.aps.org/doi/10.1103/PhysRevLett.103.150502}
}

@misc{cplex,
  author = {{IBM Corp.}},
  title = {{IBM ILOG CPLEX Optimization Studio}},
  howpublished = {\url{https://www.ibm.com/products/ilog-cplex-optimization-studio}}
}

@misc{gurobi,
  author = {{Gurobi Optimization, LLC}},
  title = {{Gurobi Optimizer Reference Manual}},
  year = 2026,
  url = "https://www.gurobi.com"
}

@book{wood2013power,
  title={Power Generation, Operation, and Control},
  author={Wood, A.J. and Wollenberg, B.F. and Shebl{\'e}, G.B.},
  isbn={9780471790556},
  lccn={2013013050},
  url={https://books.google.de/books?id=JafyAAAAQBAJ},
  year={2013},
  publisher={Wiley}
}

@article{Hoelscher_2026,
doi = {10.1088/1751-8121/ae4c31},
url = {https://doi.org/10.1088/1751-8121/ae4c31},
year = {2026},
month = {mar},
publisher = {IOP Publishing},
volume = {59},
number = {12},
pages = {125301},
author = {Hölscher, Leonhard and Karch, Lukas and Samimi, Or and Danzig, Tamuz},
title = {Quantum simulation-based optimization for cooling system design},
journal = {Journal of Physics A: Mathematical and Theoretical},
}

@article{Hoelscher_2026_2,
doi = {10.1088/2058-9565/ae4cfe},
url = {https://doi.org/10.1088/2058-9565/ae4cfe},
year = {2026},
month = {mar},
publisher = {IOP Publishing},
volume = {11},
number = {2},
pages = {025029},
author = {Hölscher, Leonhard and Ahrend, Oliver and Karch, Lukas and L’Estocq, Carlotta and Andreu, Marc Marfany and Stollenwerk, Tobias and Wilhelm, Frank K and Kowalski, Julia},
title = {End-to-end quantum algorithm for topology optimization in structural mechanics},
journal = {Quantum Science and Technology}
}

@misc{Shewchuck_ConjugateGradientMethods,
author = {Shewchuk, Jonathan R},
title = {An Introduction to the Conjugate Gradient Method Without the Agonizing Pain},
year = {1994},
publisher = {Carnegie Mellon University},
address = {USA}
}

@misc{apte2025iterativeinterpolationschedulesquantum,
      title={Iterative Interpolation Schedules for Quantum Approximate Optimization Algorithm}, 
      author={Anuj Apte and Shree Hari Sureshbabu and Ruslan Shaydulin and Sami Boulebnane and Zichang He and Dylan Herman and James Sud and Marco Pistoia},
      year={2025},
      eprint={2504.01694},
      archivePrefix={arXiv},
      primaryClass={quant-ph},
      url={https://arxiv.org/abs/2504.01694}, 
}

@article{Bucher_2025,
   title={Penalty-free approach to accelerating constrained quantum optimization},
   volume={112},
   ISSN={2469-9934},
   url={http://dx.doi.org/10.1103/fb5m-cl9m},
   DOI={10.1103/fb5m-cl9m},
   number={6},
   journal={Physical Review A},
   publisher={American Physical Society (APS)},
   author={Bucher, David and Stein, Jonas and Feld, Sebastian and Linnhoff-Popien, Claudia},
   year={2025},
   month=Dec }

\end{document}